# Persistent Laminar Flow at Reynolds Numbers Exceeding 100,000

John O. Dabiri[1,2,*], Nina Mohebbi[1], and Matthew K. Fu[1]

1. Graduate Aerospace Laboratories, California Institute of Technology, Pasadena, CA 91125, USA

2. Mechanical and Civil Engineering, California Institute of Technology, Pasadena, CA 91125, USA

**Accurate prediction of the transition from laminar flow to turbulence remains an unresolved challenge despite its importance for understanding a variety of environmental[1,2,3,4,5], biological[6,7,8,9,10], and industrial phenomena[11,12,13]. Well over a century of concerted effort has aimed toward a quantitative, mechanistic explanation of Osborne Reynolds' seminal observation of transition to turbulence in pipe flow[14,15,16,17,18,19,20,21,22], typically occurring when the ratio of inertial and viscous fluid dynamic forces—the eponymous Reynolds number—is approximately 2000. These studies have been confounded by subsequent observations[23] that the Reynolds number at which transition occurs can be delayed to values as high as 100,000. This record-high laminar Reynolds number has not been exceeded in any similar flow configuration for more than 70 years, however, as it required experiments to be conducted using pipe lengths of up to 18 meters housed within a bomb shelter to eliminate ambient disturbances to the flow. Here, we demonstrate a benchtop jet flow that exhibits persistent laminar flow beyond a Reynolds number of 116,000, a value limited only by the maximum flow-generating**

**capacity of the current facility. High-speed videography of the jet shape and flow velocimetry within the jet confirm the laminar nature of the flow, even in the presence of visible ambient flow disturbances arising from non-idealities in the facility design. The measured spatial evolution of the velocity profile within the jet, approaching a "top hat" shape with increasing downstream distance, is qualitatively distinct from the velocity profile evolution in laminar pipe flow and appears to promote persistence of the laminar flow. These results suggest the existence of an empirically accessible flow regime in which turbulence might be circumvented at arbitrarily high Reynolds numbers. Moreover, the straightforward access to this unexplored flow regime that is achieved by this experimental platform can enable new insights into the nature of laminar flow and turbulence.**

The transition from laminar flow to turbulence has commonly been studied in the context of pipe flow due to its relatively simple geometry and its historical relevance to industrial processes such as the conveyance of fluids[15,16,17,18,19,20,21,22,23,24]. Empirical studies of pipe flow observe a transition to turbulence when the Reynolds number, $Re = \overline{U}D/\nu$ (where $\overline{U}$ is the average flow speed, $D$ is the diameter of the flow, and $\nu$ is the kinematic viscosity of the fluid) exceeds a value of approximately 2000 (ref. 15). Reynolds himself succeeded in delaying the transition to turbulence to $Re \approx 13,000$ by reducing disturbances within the flow entering the pipe in his facility[14]. Subsequent investigators demonstrated further increases in the Reynolds number at which laminar flow could be maintained, culminating in the experimental campaign of Pfenninger[23], who observed laminar flow until $Re = 100,100$. Theoretical models[25] of turbulence transition have accounted for these observations by predicting that the minimum disturbance sufficient to trigger transition to turbulence scales as $1/Re$. Hence, achievement of laminar flow at the high Reynolds numbers observed by Pfenninger[23] would be possible only with careful effort to eliminate all but the very smallest ambient disturbances. Indeed, the experiments of Pfenninger were conducted in a pipe up to 18 meters in length to separate the downstream flow from inlet disturbances, and it was also housed within a bomb shelter to isolate the pipe from the flow generation apparatus and other noise sources. Even with these measures, laminar flow could not be maintained above $Re = 100,100$.

We hypothesized that a different flow configuration, specifically a vertically-oriented, axisymmetric water jet in air, could be designed to maintain laminar flow at arbitrarily high Reynolds numbers, without the need for extraordinary measures to minimize flow disturbances. This arrangement maintains the geometric simplicity of pipe flow, but it also eliminates wall friction and wall vibration as potential sources of perturbation to the flow. In place of solid walls, the air-water interface of the jet introduces a potential new source of turbulence via instabilities associated with deformation and mobility of the interface. Susceptibility of the jet to break-up by capillary instabilities is characterized by the Weber number ratio between inertial and surface tension forces, i.e., $We = \rho \bar{U}^2 D/\gamma$ (where $\rho$ is the fluid density and $\gamma$ is the surface tension at the air-water interface)[26]. Because the relative length of the jet prior to break-up, $L_b/D$, is predicted to scale as $L_b/D \sim \sqrt{We}$—provided that the jet remains laminar—the high flow speed corresponding to high Reynolds number laminar flow is predicted to increase the distance over which the jet remains coherent[27,28].

To test this prediction, we modified a nozzle designed to create an initially laminar liquid jet in air. Flow was pumped through a cylindrical canister of diameter $D_C = 11.4$ cm containing porous foam with a 5-mm nominal pore size to disrupt turbulent eddies in the water (Fig. 1). Water in the cylinder then exited the flow conditioner vertically downward through an orifice of diameter $D_O$ = 0.95 cm in the bottom endplate of the cylinder. This 12:1 flow contraction served to further straighten and accelerate the flow to the final, narrower jet diameter $D_{jet}$ = 0.84 cm. The downward-oriented jet efflux was collected in a 120-liter reservoir with its water surface located 20 cm below the nozzle orifice. A three-stage pump system comprising a pump submerged inside the reservoir, followed by two additional pumps connected in series, returned flow back to the nozzle to create a continuously recirculating flow loop. Further details of the facility design are provided in the Methods.

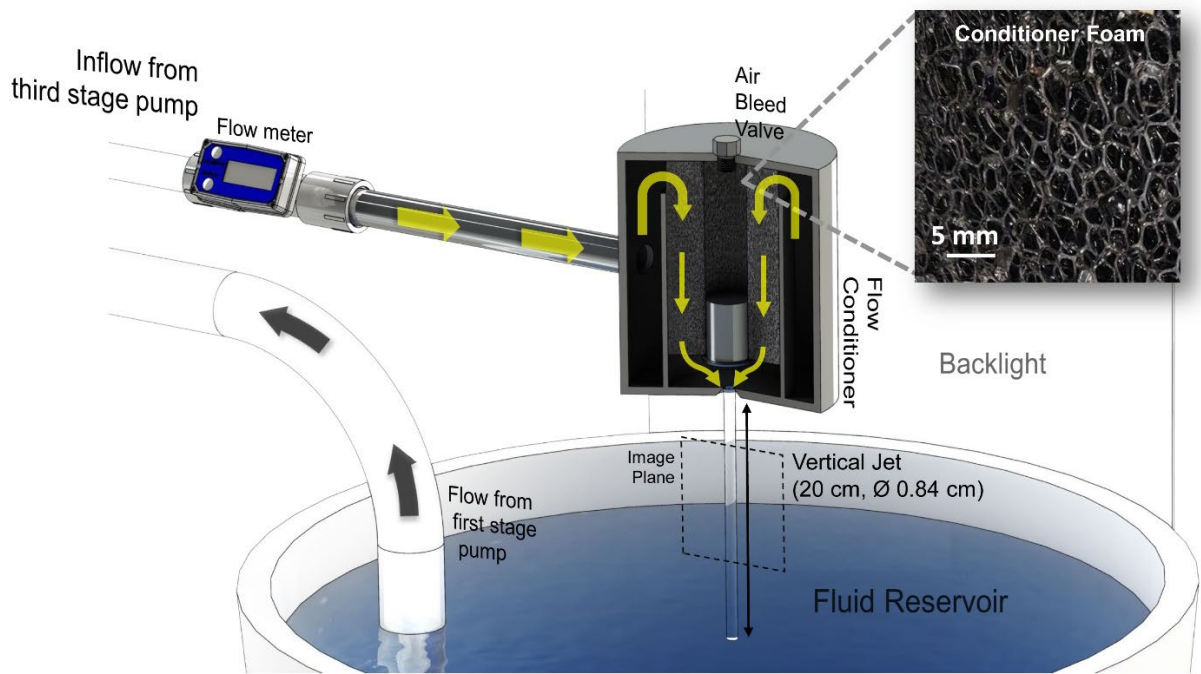

**Figure 1.** Schematic of high Reynolds number liquid jet facility. Water from fluid reservoir (bottom) is transported through three pump stages into flow meter (upper left) followed by flow conditioner (center). Yellow arrows indicate flow paths within flow conditioner, including passage through conditioner foam (detail in inset). Flow exits the flow conditioner vertically

downward. The image plane is illuminated by a backlight and recorded by a high-speed camera. Further details of the apparatus are provided in the Methods section.

The flow conditioning cylinder was modified to enable air trapped within it to be bled through a valve on its top surface when the jet was oriented vertically downward. This air evacuation was necessary to prevent air within the cylinder from disrupting the laminarizing function of the flow conditioner. Conversely, a turbulent jet could be intentionally created by closing the air bleed valve, in which case air entrained by the pump system accumulated over time within the cylinder and disrupted its normal laminarizing function.

The resulting jet was backlit and recorded using high-speed videography to characterize the flow kinematics at a spatial resolution of 1024 x 1024 pixels and at 20,000 frames per second (fps). Fig. 2 compares snapshots of the jet at Reynolds numbers ranging from 10,600 to 116,100 ($We = $ 180 to 22,100, respectively). In cases for which a turbulent jet was intentionally induced, the jet exhibits increasingly fine structure at higher Reynolds numbers, including sporadic spray formation from the lateral surfaces of the jet (e.g., Fig. 2g). By contrast, the initially laminar jets maintain an orderly flow structure even at the highest Reynolds number achievable with the available pump system, $Re = 116,100$.

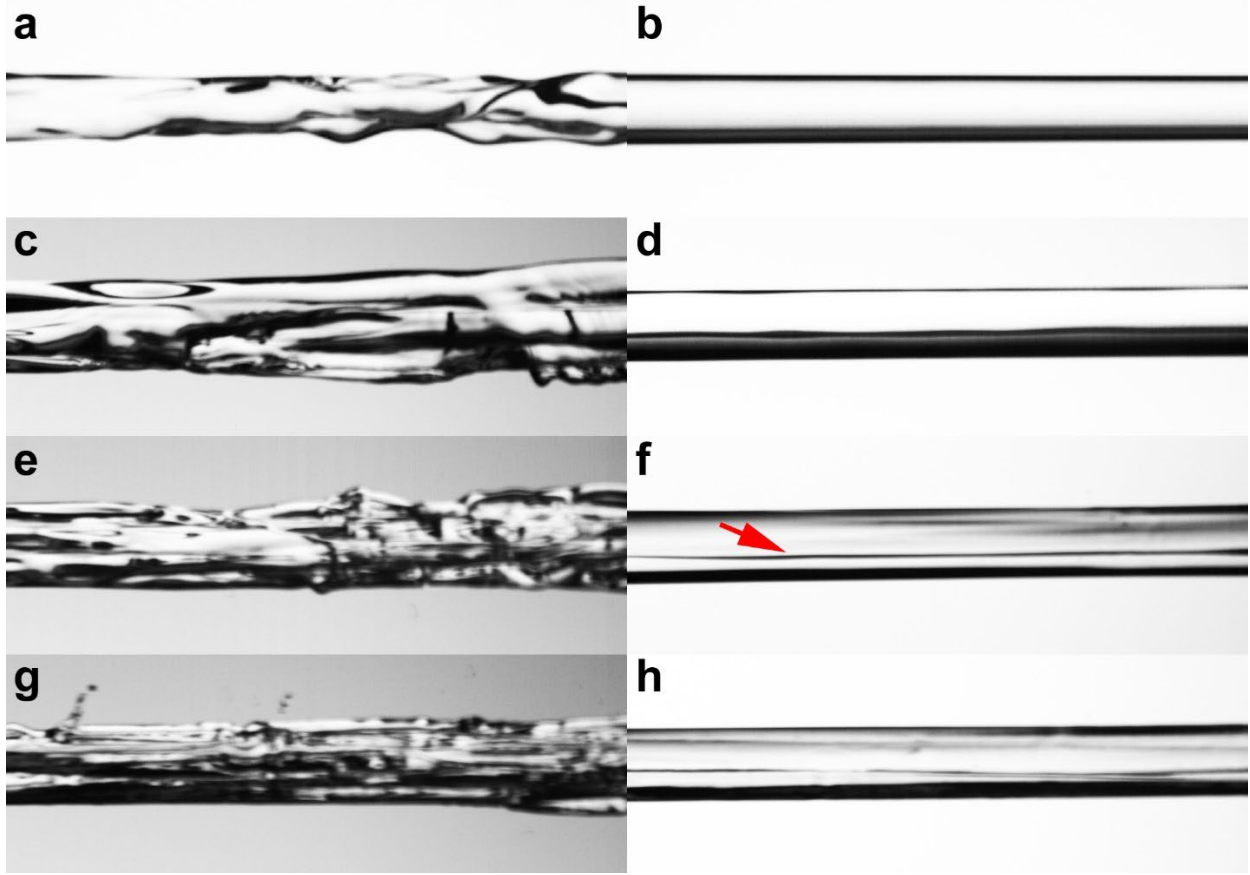

**Figure 2.** High-speed videography of turbulent and laminar jets created in the current facility. Images were collected at 20,000 fps and are rotated 90 degrees anticlockwise, so that flow is oriented from left to right. Laminar jet diameter is 0.84 cm. Left column, turbulent jets created by intentionally disrupting flow conditioner. Right column, laminar jets created during normal operation. Reynolds numbers based on mean jet speed and diameter are $Re$ = 10,600 (**a** and **b**), $Re$ = 29,800 (**c** and **d**), $Re$ = 92,100 (**e** and **f**), and $Re$ = 116,100 (**g** and **h**). Dark line visible within the laminar jets (e.g., red arrow in panel f) corresponds to a shadow cast by a necessary structural support between the light source and the jet.

High-speed videography (see Supplementary Video 1 at

https://www.its.caltech.edu/~jodabiri/largeweb/SupplementaryVideo1.mp4) revealed that the laminar flow persisted despite sporadic perturbations to the jet, e.g., due to fluctuations in the pump flow rate, the presence of air bubbles entrained into the recirculating flow loop from the reservoir, and mechanical vibrations of the entire apparatus. This robustness to non-idealities in the experiment—normally the sources of broadband convective instabilities—is in stark contrast to the observed sensitivity of laminar pipe flow at high Reynolds numbers, which necessitated the extraordinary measures taken to achieve laminar flow at these high Reynolds numbers in prior experiments[23].

While the visualizations in the right-hand column of Fig. 2 are qualitatively consistent with laminar flow, direct assessment of laminarity requires quantification of the Lagrangian trajectories of fluid particles comprising the jet[29]. We pursued this via two approaches. First, we leveraged the fact that the air-water interface of the jet is comprised of fluid particles by definition. Hence, the kinematics of the interface over time provide an indirect means to infer the presence of laminar flow. A time series of 3000 images of each jet was captured at 20,000 fps to analyze fluctuations in the air-water interface. In the highest Reynolds number case, $Re$ = 116,100, this time series corresponds to 249 jet diameters of downstream travel by the jet, i.e., $\bar{U}t/D$ = 249. Figs. 3a and 3b illustrate the interface detection applied to example images of turbulent and laminar jets, respectively, at $Re$ = 116,100. The average length of the two lateral interfaces, $L$, was compared with the axial length of the jet, $L_0$, to quantify the straightness and unsteadiness of the jet.

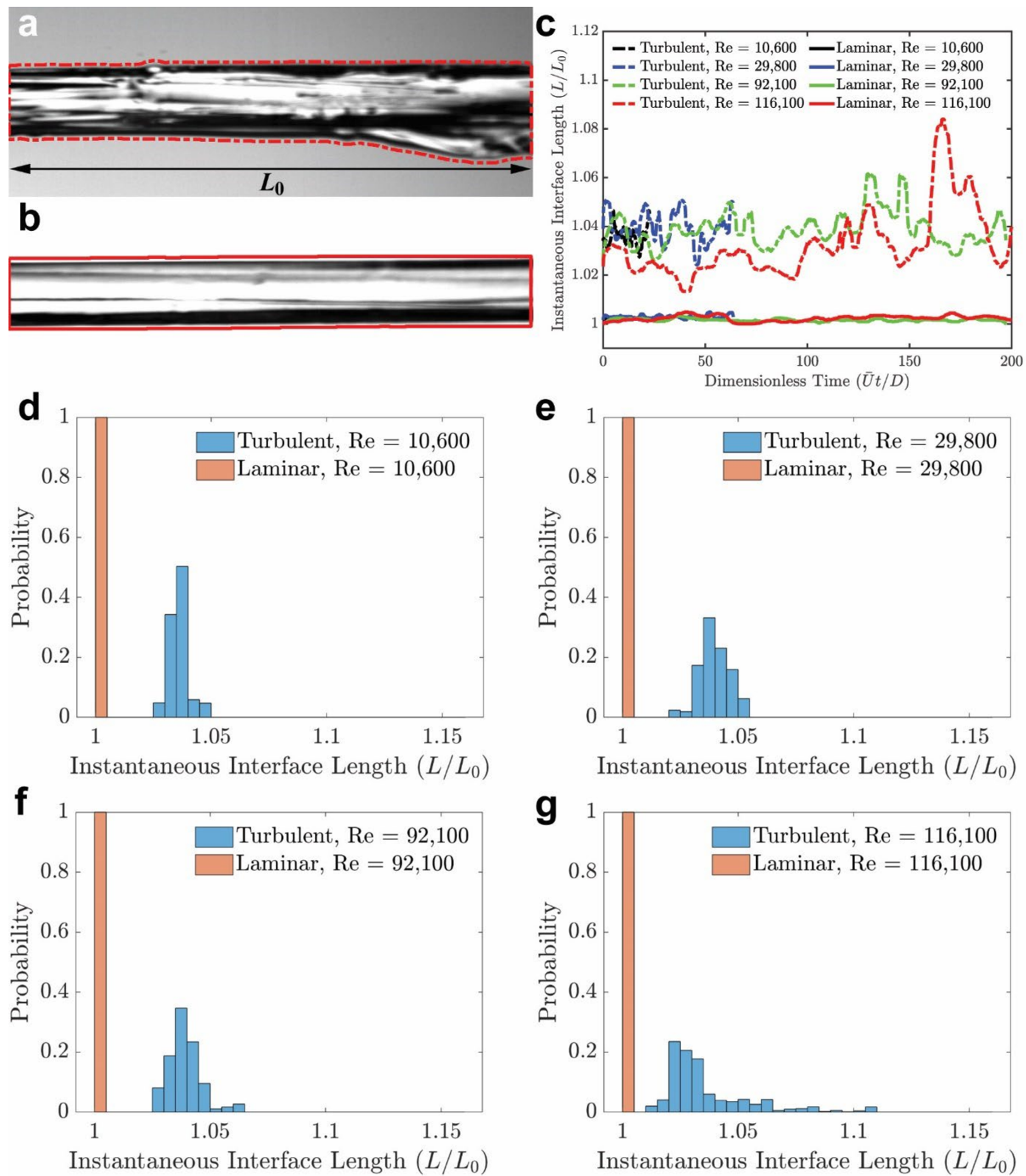

**Figure 3.** Analysis of jet interface kinematics. **a,** Turbulent jet at *Re* = 116,100. Dashed line indicates jet boundary identified by automated image analysis (see Methods). Reference jet length $L_0$ is indicated for comparison with the mean of the two lateral jet interface lengths. **b,** Laminar jet at *Re* = 116,100. Jet diameter is 0.84 cm. **c,** Instantaneous jet interface length $L/L_0$

versus dimensionless time $\overline{U}t/D$. Laminar jet cases are shown in continuous curves; turbulent jet cases shown in dashed curves. Color correspondence to Reynolds number is indicated in the legend. **d-g**, Distributions of interface length for laminar and turbulent jets at four Reynolds numbers from *Re* = 10,600 to *Re* = 116,100.

Fig. 3c plots the instantaneous interface length versus normalized time for laminar and turbulent jets at four Reynolds numbers from 10,600 to 116,100. The slower jet speed at lower Reynolds numbers leads to a correspondingly shorter normalized time, $\overline{U}t/D$, for the measurement samples from those cases. Nonetheless, a consistent distinction is observed in the behavior of the laminar and turbulent jet interfaces. Indeed, the distributions of interface lengths for the pair of laminar and turbulent flows at each Reynolds number are statistically significantly different (two-sample *t*-test, $p < 0.001$). Whereas the turbulent jet interface lengths show a broadening distribution with increasing Reynolds number, the laminar jet interface length exhibits a persistently narrow distribution close to $L/L_0 = 1$ (Figs. 3d-g). This latter observation is consistent with a jet interface comprising fluid particles that move downstream in a laminar fashion.

A more direct assessment of flow laminarity was achieved by tracking the motion of tracers advected passively in the bulk flow of the jet at *Re* = 116,100. To be sure, imaging within the bulk of the jet is only feasible when the jet is laminar, as turbulence would distort the air-water interface thereby preventing direct optical access to particles suspended in the flow. Hence, our ability to collect this type of data is further, implicit evidence of laminar flow. To achieve visual access in the present case, the backlit jets were seeded with a sparse concentration of 13-micron diameter, neutrally buoyant particles. To visualize tracer particles in a region spanning the majority of one half of the jet, i.e., from the centerline to the lateral air-water interface, it was necessary to offset the light source from the camera optical axis in such a way that light refraction primarily obscured the opposite lateral interface. A necessary structural support for the facility created an additional axial shadow that prevented tracking of some particles in that region of the jet (cf. Fig. 2f).

Despite these limitations, the imaging method was successful in tracking particles as they were advected within the bulk of the jet. Figure 4a shows a snapshot of the flow recorded at 20,000 fps. In addition to the 13-micron diameter particles introduced into the flow loop, larger particles were also present in the flow at sparse concentrations, either due to agglomeration of the tracer particles or from external particles entering the flow from the reservoir (e.g., red arrow in Fig. 4a). These particles were excluded from subsequent analysis. The responsiveness of the seeded particles to flow fluctuations can be quantified by using the Stokes number, $Sk = (\rho_p d_p^2 u')/\mu D$, where $\rho_p$ and $d_p$ are the density and diameter, respectively, of the tracer particles, $\mu$ is the dynamic viscosity of water, $u'$ is magnitude of velocity fluctuations in any direction, and $D$ is a characteristic length scale, taken here as the jet diameter[30]. The Stokes number of the tracer particles used presently remains in the range $Sk < 0.1$ for turbulence fluctuations of up to 30 percent of the mean jet speed ($\bar{U}$ = 13.9 m/s), a turbulence intensity that is double the maximum values typically encountered in smooth-wall pipe flow[31,32]. Therefore, the tracer particles used here can faithfully track turbulence in the flow if it is present.

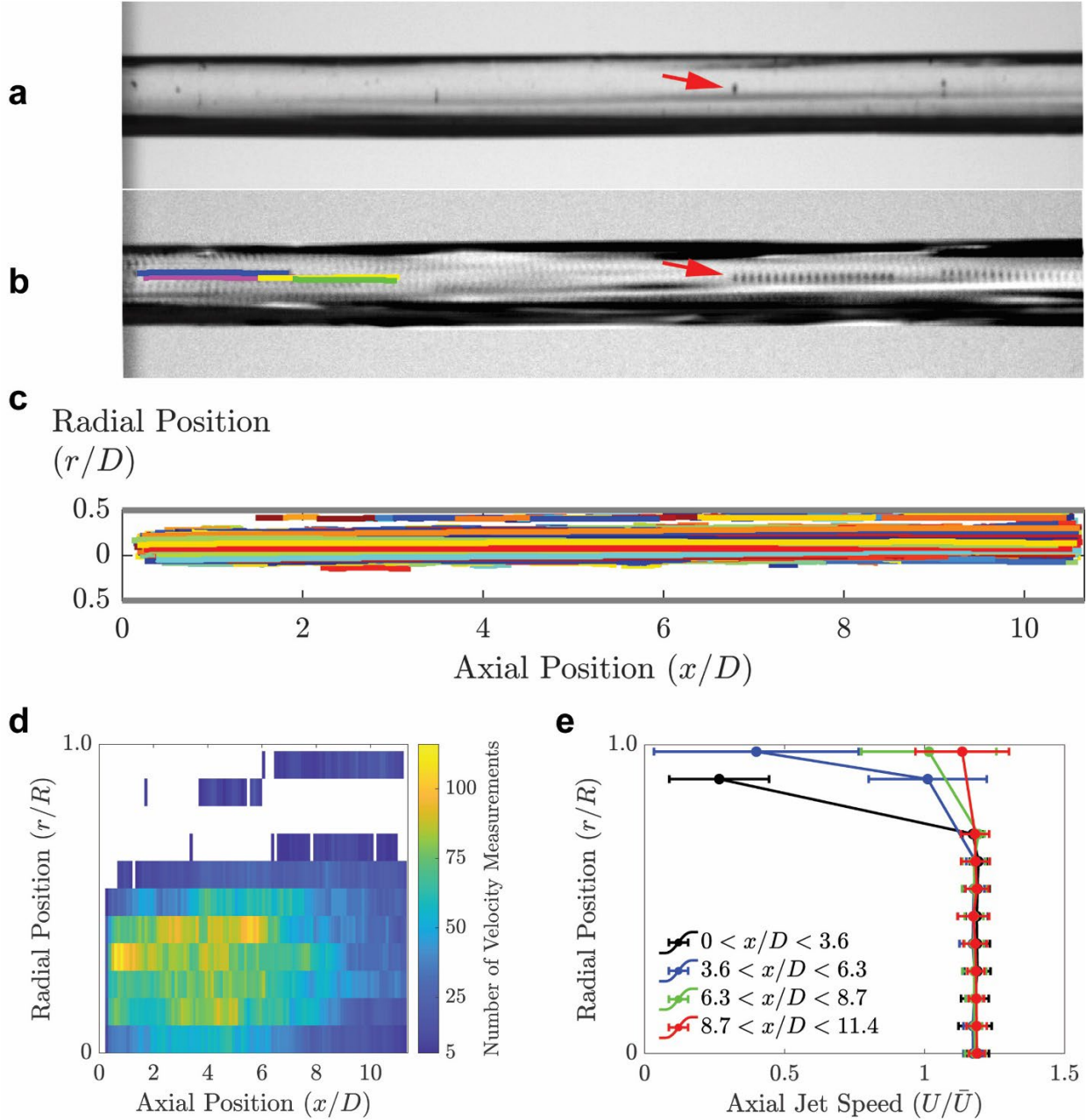

**Figure 4.** Particle tracking velocimetry measurements of the laminar jet at $Re$ = 116,100. **a,** Example snapshot of the jet seeded with 13-micron tracer particles. Additional, larger particles were observed in the flow (e.g., red arrow) but were excluded from the subsequent analysis. Jet diameter is 0.84 cm. **b,** Time series of 20 overlaid images, illustrating particle displacement during the 50 microseconds between recorded images. Colored lines show particle trajectories

reconstructed from particle positions (see Methods). Red arrow identifies particle track excluded from analysis. **c,** Plot of 2403 particle trajectories colored randomly. Trajectories correspond to measurements from three recordings sampled several minutes apart. **d,** Spatial distribution of velocity measurements within jet. Color corresponds to number of measurements in each rectangular region. Number of measurements ranges from 5 (minimum allowed threshold) to 116. **e,** Spatially-averaged velocity profiles at successive downstream locations along the jet. The reference value $\bar{U}$ is computed using the volume flow rate $Q$ measured by the flowmeter and the visually measured jet diameter, i.e., $\bar{U} = 4Q/(\pi D_{jet}^2)$.

The temporal resolution of the images was sufficient to reconstruct tracer particle trajectories along the full visible length of the jet, an axial distance of more than 11 jet diameters. Fig. 4b plots a sample of the trajectory reconstruction overlaid on a sequence of 20 images, showing the particle displacement between frames. A total of 2403 trajectories were recorded to characterize the laminar jet at $Re$ = 116,100 (Fig. 4c). These trajectories, derived from three separate video samples recorded over several minutes, are illustrative of persistent laminar flow within the bulk of the jet (see also Supplementary Video 2 at https://www.its.caltech.edu/~jodabiri/largeweb/SupplementaryVideo2.mp4). This result is consistent with the conclusions derived from the interface analysis in Fig. 3.

The 2403 tracer particle trajectories comprise 38,644 individual velocity measurements corresponding to the particle displacements during each 50-microsecond interval of the recorded flows. Fig. 4d shows the spatial distribution of those measurements, where we have required a minimum of 5 velocity measurements in order for a given region to be considered properly sampled. As many as 116 measurements are recorded in locations where several tracer particles were advected during the sampling periods. Because this imaging technique is a planar projection of tracer particles being advected in a three-dimensional, cylindrical volume, measurements closer to axis of symmetry represent an amalgam of particles spanning the full depth of the jet, both at the axis of symmetry and also at the air-water interfaces located in front and behind the axis of symmetry relative to the camera position. This limitation will be revisited momentarily in the discussion of the velocity measurements.

The region that is absent of any velocity measurements, primarily near the air-water interface (i.e., $r/R = 1$) is a consequence of light refraction away from the air-water interface and the aforementioned shadowing from the structural support.

Cross-sectional averages of the velocity at successive streamwise locations reveal an evolution of the velocity profile of the laminar jet as it propagates downstream (Fig. 4e). Specifically, the flow emerging from the nozzle orifice is faster along the centerline of the jet than at its lateral edges. As the jet moves downstream, the flow near the air-water interface accelerates, causing the velocity profile to approach a "top-hat" shape by the time it reaches the downstream extent of the measurement domain at $x/D = 11.4$.

To be sure, the planar projection effect described above can bias velocity measurements closer to the axis of symmetry to lower values, as slower-moving particles near air-water interfaces in front and behind the axis of symmetry will be included in the measurement averages in those locations. This effect could potentially lead to underestimation of the jet velocity near the centerline.

Additional measurement uncertainty is associated with the precision of the visual measurement of the jet diameter used to calculate $\overline{U}$. This uncertainty, combined with the limited number of measurement sites near the lateral edges of the jet, are likely sources of the apparent discrepancy between a spatial average of the velocity profiles determined from particle tracking and the mean jet speed $\overline{U}$ calculated based on the volume flow rate and the measured jet diameter (i.e., a spatially averaged value of $U/\overline{U} \neq 1$ in Fig. 4e).

Flow acceleration along the lateral edge of the jet is clearly observed despite the aforementioned measurement limitations. The estimated magnitude of spatial flow acceleration based on the outermost velocity measurements, $du/dx \approx 140$ s$^{-1}$, is orders-of-magnitude higher than what could be attributed to viscous diffusion, which has a time scale of order $\nu/\delta^2 \approx 1$ s$^{-1}$ for a lateral diffusion distance $\delta \approx 1$ mm across the jet. In contrast, the observed flow acceleration is consistent with the known dynamics of the *vena contracta* formed by flow forced through a narrow orifice. In such flows, a pressure gradient is established along the lateral edges of the jet where it narrows from the larger orifice diameter, which can accelerate the local flow[33]. In the present case, the jet cross-sectional area contracts near the *vena contracta* by approximately 22% from $D_O = 0.95$ cm at the orifice to $D_{jet} = 0.84$ cm in the jet within a downstream distance of less than one jet diameter. The corresponding spatial acceleration $(\overline{U_{jet}} - \overline{U_O})/D_{jet} \approx 377$ s$^{-1}$ is consistent with the magnitude of observed velocity profile evolution in the downstream jet.

Extrapolation of the evolving velocity profile suggests the possibility of an asymptotic jet shape with a nearly uniform velocity profile (excluding edge effects due to shear exerted by the surrounding air). Such a jet shape would lack the velocity gradients necessary to induce turbulence at any Reynolds number. To the contrary, time resolved measurements of the laminar jet at $Re$ = 116,100 illustrate the apparent suppression of perturbations as the jet surface accelerates (Supplementary Video 1). This observation is consistent with prior studies of liquid jets in the shear breakup regime, which also exhibit robustness to inlet perturbations[26,34]. Moreover, while the effect of spatial velocity profile evolution on the stability of liquid jets has been the subject of theoretical debate[35,36], a prior empirical study reported a similar correlation between initially irrotational laminar flow and complete suppression of subsequent liquid jet break-up, albeit with large-scale, sinuous and helical waves on that jet at $x/D > 30$ (ref. 37). Coherence of the jets created in the present facility could be maintained further downstream by a similar mechanism.

The air-water interface of the jet experiences a non-zero shear due to the inertia of the surrounding air, which could in principle affect the stability of the jet. However, the magnitude of that shear stress is proportional to the density of air, which is three orders-of-magnitude smaller than the water density. It is therefore unlikely to be a significant factor in modifying the velocity profile of the jet provided that the interface remains aligned parallel to the jet axis. Any associated perturbation to the jet could be mitigated by conducting experiments in vacuum conditions (within limits that avoid phase transition).

A more immediate and fascinating question to explore is whether there exists any limit on the Reynolds numbers at which this flow can remain laminar. The present facility was limited only by the maximum flow rate achievable by the available pumps. There appear to be no conceptual constraints on the achievement of persistent laminar flow at even higher Reynolds numbers than those demonstrated here, a conclusion that can inspire new advances in our understanding of turbulence. For example, the lack of turbulence in the present case is actually consistent with the prediction of linear stability theory for pipe flows with similar velocity gradient profiles[15,38]. The fact that pipe flows are empirically observed to transition to turbulence, in contradiction to the theoretical prediction, has been previously reconciled by appealing to nonlinear flow phenomena[15,39]. A persistent laminar flow in the current configuration could instead call into question the only substantive distinction between theoretical models of the present flow and spatially-evolving pipe flows, namely, representation of solid walls via the so-called "no-slip" condition[40]. That ansatz is arguably the only component of the Navier-Stokes model of fluid flow that cannot be derived from first principles[41,42,43]. The present results may indirectly challenge the ability of the no-slip condition to properly capture the role of solid walls in turbulence transition.

**Methods**

Flow facility

Flow was generated by three pumps connected in series. A 9.9-L/s maximum capacity (i.e., at zero head) asynchronous submersible magnetic drive pump (EasyPro EPA9500) was located in the 120-liter reservoir and transferred water to a second, 3.4-L/s maximum capacity pump (EasyPro EP3200N) that was connect by 5.1-cm diameter Tygon tubing. This second pump transferred water via 2.5-cm diameter Tygon tubing to a third, 1.1 L/s maximum capacity pump (EasyPro EP1050). The third pump transferred water through 2.5-cm diameter Tygon tubing to a turbine flowmeter (Omega TM10NQ9GMA) with a measurement range of 0.32 L/s to 3.2 L/s and a measurement accuracy of ±3%. The outlet of the flowmeter was connected via a 90-degree polyvinyl chloride (PVC) coupler to the inlet of the flow conditioner nozzle (Fig. 1). The flow conditioner nozzle was modified from a commercially-available laminar stream fountain (Underwater Warehouse ELN75) to enable operation in a vertically downward orientation by bleeding trapped air from the rear power cable port. Flow exiting the nozzle traveled approximately 20 cm vertically downward before being collected at the free surface of the 120-liter reservoir, completing the flow loop.

Flow visualization

The jet flow was illuminated from the rear using a rectangular, planar floodlight. For visualizations of the jet in Figs. 2 and 3 and Supplementary Video 1, the floodlight was approximately centered on the jet and oriented toward the optical axis of the camera. For particle tracking measurements (i.e., Fig. 4 and Supplementary Video 2), the floodlight was offset from the camera optical axis to minimize dark regions due to light refraction at one of the lateral air-water interfaces. For particle tracking velocimetry, the water was seeded sparsely with 13-micron diameter silver-coated, hollow glass spheres (Potters Industries, Conduct-O-Fil, SH400S20).

Flow in each case was recorded using a high-speed camera (Photron FASTCAM SA-Z) at 20,000 fps. The exposure time was set to either 20 or 50 microseconds, depending on which setting provided clearer images of the jet. Measurements at Reynolds numbers above 10,600 in Figs. 2 and 3 and Supplementary Video 1 were collected using a telephoto zoom lens (Ruili 420-800mm F/8.3-16). All measurements at $Re$ = 10,600 and particle tracking measurements in Fig. 4 and Supplementary Video 2 were collected using a combination of a 1.7X teleconverter (Nikon AF-S TC-17E II) and a telephoto zoom lens (Nikon AF-S 70-200MM F/2.8E). This latter optical setup facilitated sharper particle images for particle tracking velocimetry.

To avoid inadvertent splashing of the camera lens by the turbulent jets, the camera was rotated 90 degrees in a horizontal plane from the jet location, and a glass mirror was used to reflect the image of the jet onto the camera sensor.

Image analysis

For preparation of Fig. 2, selected images were imported into Adobe Photoshop to increase the image brightness (+100) and contrast (+50). Images collected with the shorter exposure time of 20 microseconds were brightened by an additional +100 and contrasted by an additional +50.

All other image analysis was performed using Matlab (The MathWorks, Inc.). Interface analysis was accomplished using Matlab functions for edge detection and manipulation of binary images (Code available upon request). Particle tracking was similarly accomplished using thresholding from grayscale images to binary. Larger particles were excluded from the analysis (cf. Figs. 4ab) by enforcing a maximum image size criterion of 32 pixels per particle. Some tracer particles appear larger when imaged by backlighting due to blurring and lensing from the non-planer air-water interfaces. Hence, the image analysis is conservative in potentially excluding particles that are physically small but appear larger.

Due to the sparse concentration of particles in the flow, a nearest-neighbor search between successive frames, combined with search criteria that required particles to move downstream and by amounts limited by the order of magnitude of the flow speed, was sufficient to match particles between frames (Code available upon request). The convex hull of the series of particles comprising each trajectory was used to qualitatively confirm each trajectory (see Supplementary Video 2). The centerline of each convex hull, extracted using the "bwskel" function in Matlab, was used to quantify particle trajectories in Fig. 4.

Data availability

A high-resolution version of Supplementary Video 1 will be deposited at the Caltech Research Data Repository (https://data.caltech.edu).

**Acknowledgements**

The authors gratefully acknowledge support from Federico Rios Tascon in testing of an early prototype jet apparatus; and Robert Whittlesey, Clara O'Farrell, and Daniel Araya for discussion of the results. The project was supported by funding from the National Science Foundation Alan T. Waterman Award.

**Author contributions**

J.O.D. conceived of the project and conducted the experiments reported in this manuscript. N.M. conducted experiments on earlier prototype designs that informed the final experiment design. M.K.F. contributed data analysis. J.O.D. and M.K.F prepared the manuscript, and all authors contributed to editing of the manuscript.

**Competing interests**

The authors declare no competing interests.

**Materials & Correspondence**

Requests and correspondence should be addressed to J.O.D. (jodabiri@caltech.edu).

**Supplementary information**

Supplementary Video 1. Video of each jet condition shown in Fig. 2. Playback is 1/333 of real time.

Supplementary Video 2. Video of particle trajectories included in Fig. 4c. Particle trajectories in video are visualized using the convex hull of pixels associated with each particle image in a given trajectory. Playback is 1/333 of real time.